\documentclass[useAMS,usenatbib]{mn2e}
\usepackage{epsf}
\usepackage{amssymb}
\usepackage[usenames]{color}

\def\plotone#1{\centering \leavevmode
\epsfxsize=\columnwidth \epsfbox{#1}}
\def\plottwo#1#2{\centering \leavevmode
\epsfxsize=.99\columnwidth \epsfbox{#1} \hfil
\epsfxsize=.99\columnwidth \epsfbox{#2}}
\def\plotwide#1{\centering \leavevmode
\epsfxsize=1.4\columnwidth \epsfbox{#1}}

\def\kms{\,{\rm km~s}^{-1}}

\newcommand{\be}{\begin{equation}}
\newcommand{\ee}{\end{equation}}

\def\disp {\displaystyle}

\title[Simple recipe for estimating galaxy masses]{Testing a simple recipe for estimating galaxy masses from minimal observational data.}
\author[Lyskova et al.]{N.~Lyskova,$^{1,2}$ 
E.~Churazov,$^{1,2}$ I.~Zhuravleva,$^{1}$ 
T.~Naab,$^{1}$ L.~Oser,$^{1,3}$ O.~Gerhard,$^{4}$ 
\newauthor 
X.~Wu$^{4}$  
\newauthor \\
$^1$ Max-Planck-Institut f\"ur Astrophysik, Karl-Schwarzschild-Strasse 1, 85741
Garching, Germany\\
$^2$ Space Research Institute (IKI), Profsoyuznaya 84/32, Moscow 117810, 
Russia\\
$^3$ Universit\"ats-Sternwarte M\"unchen, Scheinerstr. 1, D-81679 M\"unchen, Germany \\
$^4$ MPI f\"{u}r Extraterrestrische Physik, P.O.\ Box 1603, 85740
Garching, Germany\\
}

\begin{document}

\pagerange{\pageref{firstpage}--\pageref{lastpage}}
\pubyear{2011}

\maketitle

\label{firstpage}
\begin{abstract}
  
The accuracy and robustness of a simple method to estimate the total mass profile of a galaxy is tested using a sample of 65 cosmological zoom-simulations of individual galaxies. The method only requires
information on the optical surface brightness and
the projected velocity dispersion profiles and therefore can be
applied even in case of poor observational data.
In the simulated sample massive galaxies ($\sigma \simeq  200-400$  $\kms$) at redshift $z=0$ have almost
isothermal rotation curves for broad range of radii (RMS $\simeq 5\%$ for the circular speed deviations from a constant value over $0.5R_{\rm eff} < r < 3R_{\rm eff}$).
For such galaxies the method recovers the unbiased value of the circular speed. The sample averaged deviation from the true circular speed is less than $\sim 1\%$  with the
scatter of  $\simeq 5-8\%$ (RMS) up to $R \simeq 5R_{\rm eff}$. 
Circular speed estimates of massive non-rotating simulated galaxies at higher redshifts ($z=1$ and $z=2$) are also almost unbiased and with the same scatter.
For the least massive galaxies in the sample ($\sigma < 150$  $\kms$) at $z=0$ the RMS deviation is $\simeq 7-9\%$ and the mean deviation is biased low by about $1-2\%$. 
We also derive the circular velocity profile from the
hydrostatic equilibrium (HE) equation for hot gas in the simulated galaxies. The
accuracy of this estimate is about RMS $\simeq 4-5\%$ for massive objects ($M > 6.5\times 10^{12} M_\odot$)
and the HE estimate is biased low by $\simeq 3-4\%$, which can be traced to the presence of gas motions.
This implies that the simple mass estimate can be used to determine the mass of 
observed massive elliptical galaxies to an accuracy of $5-8 \%$ and can be very useful for galaxy surveys.

\end{abstract}


%

\sloppypar

\section{Introduction}

The accurate determination of galaxy masses is a crucial issue for galaxy formation and evolution models. Disentangling dark matter and baryonic matter of a galaxy permits testing the predictions of $\Lambda$CDM-cosmology and probing the mass function. An algorithm for deriving the mass of a spiral galaxy is straight forward - one just need to measure a rotation curve from gas or stars that can be safely assumed to be on circular orbits. For elliptical galaxies the situation is less simple. There is no `perfect' (in terms of accuracy) tracer to measure the total gravitational potential. The main problem is the degeneracy between the anisotropy of stellar orbits and the mass. The shape of stellar orbits is not known a priory and different combinations of orbits may give the same distribution of light. Several different approaches for mass determination were proposed and succesfully implemented, like strong and weak lensing \citep[e.g.][]{2007ApJ...667..176G,2006MNRAS.368..715M}, modelling of X-ray emission of hot gas in galaxies  \citep[e.g.][]{2006ApJ...646..899H, 2008MNRAS.388.1062C}, Schwarzschild modelling of stellar orbits, etc.  Accurate data on the projected line-of-sight velocity distribution with information on higher-order moments enables an accurate determination of the mass distribution for nearby ellipticals \citep[e.g.][]{1998MNRAS.295..197G, 2011MNRAS.415..545T}.  However, in case of minimal available data detailed modelling is often not possible. Therefore it is important to find a method to measure galaxy masses with reasonable accuracy which gives an unbiased estimate when averaged over a large number of galaxies. In particular, it can be extremely useful while analysing large surveys, especially at high redshifts when detailed observational data of each individual galaxies are often not available.

The simplest way of estimating the mass of a galaxy is based on the projected velocity dispersion in a fixed aperture \citep[e.g.][]{2006MNRAS.366.1126C}. A slightly more complicated approach is described in \cite{2010MNRAS.404.1165C}. To estimate the mass the only information required is the light profile and either the dispersion profile  measurement or at least a reliable dispersion measurement at some radius.  Testing this particular method on a sample of simulated galaxies is the subject of this paper. The main questions that we want to address are 
(i) What is the accuracy of this method?  
(ii) Does it give an unbiased result? 
(iii) What are the  restrictions for application of this method?

The structure of the paper is as follows. In section~\ref{sec:method}, we provide a brief description of the method. In section~\ref{sec:sample} we describe the sample of simulated galaxies which is used to test the method. The analysis of the accuracy of the method is presented in section~\ref{sec:analysis} where we also discuss alternative methods for determining the circular velocity. A summary on the bias and accuracy of the various methods is given in section~\ref{sec:discuss} with conclusions in section~\ref{sec:conc}.

\section{Description of the method}

\label{sec:method}

The main idea of the method is described in  \cite{2010MNRAS.404.1165C}. Here we just provide a brief summary.

The method is based on the stationary non-streaming spherical Jeans equation:
\be
{d\over dr}j\sigma_r^2+2\frac{\beta}{r}j\sigma_r^2=-j{d\Phi\over dr},
\ee

where $j(r)$\footnote{Throughout this paper we denote a projected 2D radius as R and a 3D radius as r.} is the stellar luminosity density, $\sigma_r(r)$ is the radial component of the velocity dispersion tensor (weighted by luminosity), $\beta(r)~=1~-~\sigma_{\theta}^2/\sigma_{r}^2$ is the stellar anisotropy parameter  ($\sigma_{\theta}=\sigma_{\phi}$ because of the assumed spherical symmetry) and $\Phi(r)$ is the gravitational potential of a galaxy. 

While the stellar luminosity density  $j(r)$ and radial dispersion $\sigma_r(r)$ can not be observed directly they contribute to the  two-dimensional surface brightness $I(R)$ and the velocity dispersion $\sigma(R)$ profiles:

\be
I(R)=2\int_R^\infty\!\! {j(r)r\,dr\over\sqrt{r^2-R^2}},
\ee

\be
\sigma^2(R)\cdot I(R)=2\int_R^\infty\!\! j(r)\sigma_r^2(r)\left(1-\frac{R^2}{r^2}\beta(r)\right){r\,dr\over\sqrt{r^2-R^2}}.
\ee

Assuming $\beta(r)=\rm const$ we note that $\beta=0$ for systems where the distribution of stellar orbits is isotropic, $\beta=1$ if all stellar orbits are radial and $\beta \rightarrow -\infty$ if the orbits are circular.

Assuming the logarithmic form of the gravitational potential $\Phi(r)=V_c^2\ln(r)+\rm const$ and using local properties of given $I(R)$ and $\sigma(R)$ one can calculate a circular velocity $V_c$ for three different types of stellar orbits: isotropic ($\sigma_r=\sigma_{\phi}=\sigma_{\theta}$, $\beta=0$), radial ($\sigma_{\phi}=\sigma_{\theta}=0$, $\beta=1$) and circular ($\sigma_r=0$, $\beta\rightarrow -\infty$). These relations are given by: 

\[
V^{\rm iso}_c=\sigma_{\rm iso}(R) \cdot \sqrt{1+\alpha+\gamma} 
\]
\be
V^{\rm circ}_c=\sigma_{\rm circ}(R) \cdot \sqrt{2 \frac{1+\alpha+\gamma}{\alpha}}
\label{eq:main} 
\ee
\[
V^{\rm rad}_c=\sigma_{\rm rad}(R) \cdot \sqrt{\left(\alpha+\gamma\right
  )^2+\delta-1}, 
\]
where 
\be
\alpha\equiv-\frac{d\ln I(R)}{d\ln R}, \ \ \gamma\equiv -\frac{d\ln
  \sigma^2}{d\ln R},\ \ \delta\equiv \frac{d^2\ln[I(R)\sigma^2]}{d
  (\ln R)^2}.
\label{eq:agd}
\ee

In case of noisy data on the dispersion velocity profile the subdominant terms $\gamma$ and $\delta$ can be neglected, i.e. the dispersion profile is assumed to be flat, and equations (\ref{eq:main}) are simplified to:

\[
V^{\rm iso}_c=\sigma_{\rm iso}(R) \cdot \sqrt{\alpha+1} 
\]
\be
V^{\rm circ}_c=\sigma_{\rm circ}(R) \cdot  \sqrt{2\frac{\alpha+1}{\alpha}}
\label{eq:agd_simple} 
\ee
\[
V^{\rm rad}_c=\sigma_{\rm rad}(R)  \cdot  \sqrt{\alpha^2-1}. \\
\]

Let us call a sweet spot the radius at which all three curves $V^{\rm iso}_c(R), V^{\rm circ}_c(R)$ and $V^{\rm rad}_c(R)$ are very close to each other. One can hope that at the sweet spot the sensitivity of the method to the stellar anisotropy parameter $\beta$ is minimal and the estimation of the circular speed at this particular point is reasonable. E.g. from equations (\ref{eq:agd_simple}) it is clear that in case of the power-law surface brightness profile with $\alpha = 2$ and $\beta= \rm const$ the relation between the circular speed and the projected velocity dispersion does not depend on the anisotropy parameter \citep[e.g.][]{1993MNRAS.265..213G}. While the derivation of equations (\ref{eq:main}), (\ref{eq:agd_simple}) relies on the assumption about a flat circular velocity profile, tests on model galaxies with non-logarithmic potentials, non-power law behaviour of the surface brightness and line-of-sight velocity dispersion profiles and with the anisotropy parameter $\beta$ varying with radius \citep{2010MNRAS.404.1165C} have shown that the circular speed can still be recovered to a reasonable accuracy. Now we extend these tests to a sample of simulated elliptical galaxies.

This method for evaluating the circular speed is not only simple and fast in implementation but it also does not require any assumptions on the radial distribution of anisotropy  $\beta(r)$ and mass $M(r)$. 

The mathematical derivation of equations (\ref{eq:main}-\ref{eq:agd_simple}) can be found in  \cite{2010MNRAS.404.1165C}. A similar approach and analytic formulae for kinematic deprojection and mass inversion also can also be found in \cite{2010MNRAS.406.1220W} and \cite{2010MNRAS.401.2433M}.

\section{The sample of simulated galaxies}
\label{sec:sample}

\subsection{Description of the sample}
\label{subsec:sample_descript}

Simulations provide a useful opportunity to test different methods and procedures as all intrinsic properties of a system at hand are known. 
The main drawback of simulated objects is that they may not include all physical processes that take place in reality and thus may not reflect all complexity of nature. 
To test the procedure under consideration we have used a sample of 65 cosmological zoom simulations partly presented in  \cite{2010ApJ...725.23120}. 
These SPH simulations include feedback from supernovae type II, a uniform UV-background radiation field, star formation and radiative Hydrogen and Helium cooling but do not include  ejective feedback in the form of supernovae driven winds. 
Present-day stellar masses of simulated galaxies range from $2.18\times10^{10} M_{\odot} h^{-1}$ to $28.68\times 10^{10} M_{\odot} h^{-1}$ inside 30 kpc. 
The softening length used in simulations is about $R_{soft}$=400 ${\rm pc}~h^{-1}, h=0.72$. Typically the softening can affect profiles up to $\sim 3R_{soft}$, which is $\simeq 1.7$ kpc in our case. We have followed a conservative approach and restricted the analysis to radii larger than 3 kpc.
It should be noted that low-mass simulated galaxies may have no real counterparts possibly due to lack of important physical processes (e.g., significant winds) in simulations. 
However, it has been demonstrated in \cite{2011arXiv1106.5490O} that the massive simulated galaxies have properties very similar to observed early-type galaxies (see also Figure 4), i.e. they follow the observed scaling relations and their evolution with redshift.
For detailed description of simulations and included physics see \cite{2010ApJ...725.23120}.

To effectively increase the number of galaxies we have considered three independent projections of each galaxy. 
So the whole sample of simulated galaxies consists of 195 objects\footnote{Nevertheless, for calculating an error in a bias estimation (= RMS $/\sqrt{N}$) we conservately use the number of galaxies rather than the number of projections as the subsamples corresponding to different projections are not entirely independent.}.

\subsection{Isothermality of potentials in massive galaxies}
\label{subsec:iso}

\begin{figure}
\plotone{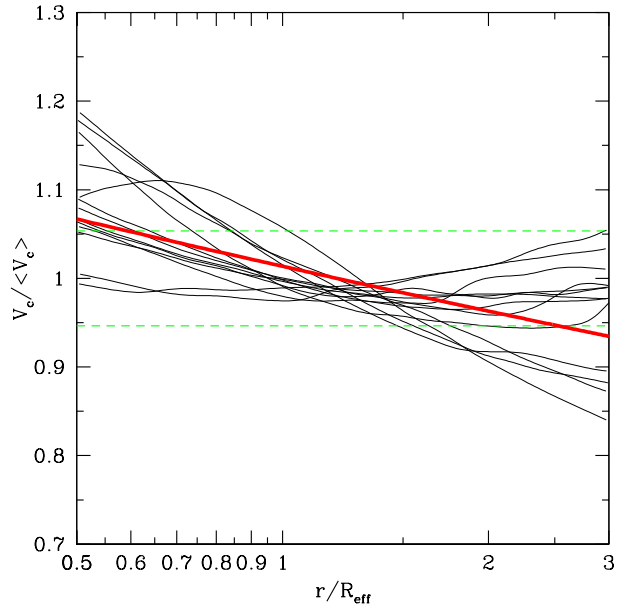}
\caption{Circular velocity curves of massive galaxies ($\sigma(R_{\rm eff})>200$ $\kms$) as a function of radius $r$.
Individual rotation curves normalised to the speed averaged over $[0.5R_{\rm eff}, 3R_{\rm eff}]$ are shown in black, green dashed lines indicate the interval $[1-RMS,1+RMS]$, where $RMS=4.9\%$, the red thick line represents the overall trend $V_c \propto r^{-0.06} $.
\label{fig:isothermal}
}
\end{figure}

First of all we have found that massive galaxies in the sample have almost isothermal
 rotation curves over broad range of radii. To demostrate this statement (Figure \ref{fig:isothermal}) we have selected galaxies with a projected velocity dispersion at the effective radius $\sigma(R_{\rm eff})$ (procedure of computation $R_{\rm eff}$ is described in section \ref{subsec:analysis}) greater than 200 $\kms$ and plotted their circular velocity curves $V_c=\sqrt{GM(<r)/r}$ as a function of $r/R_{\rm eff}$. $G$ is the gravitational constant, $M(<r)$ is the mass enclosed within $r$ and $R_{\rm eff}$ is the effective radius of the galaxy. The circular velocity curves were normalised to the value of $V_c$ averaged over $r  \in [0.5R_{\rm eff}, 3R_{\rm eff}]$. Three circular velocity curves that make the most significant contribution to the RMS actually correspond to galaxies with the effective radius $R_{\rm eff} < 6$ kpc. The fact that for these galaxies $0.5R_{\rm eff}$ is close to the softening length may affect the scatter.

\subsection{Analysis procedure}
\label{subsec:analysis}	

The analysis of each galaxy consists of several steps described below.

\begin{figure*}
\plotwide{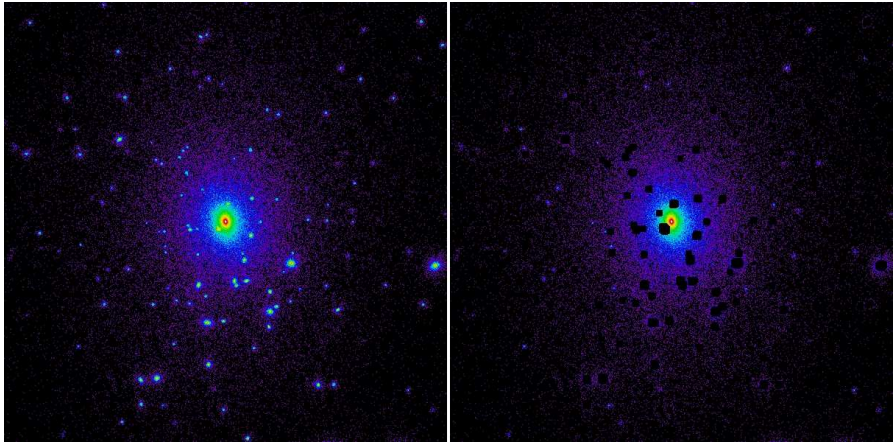}
\caption{Excluding the satellites. 150 kpc $\times$ 150 kpc. {Left:} Initial galaxy image.
{Right:} Cleaned galaxy image.  
\label{fig:M0053}
}
\end{figure*}
\begin{figure*}
\plotone{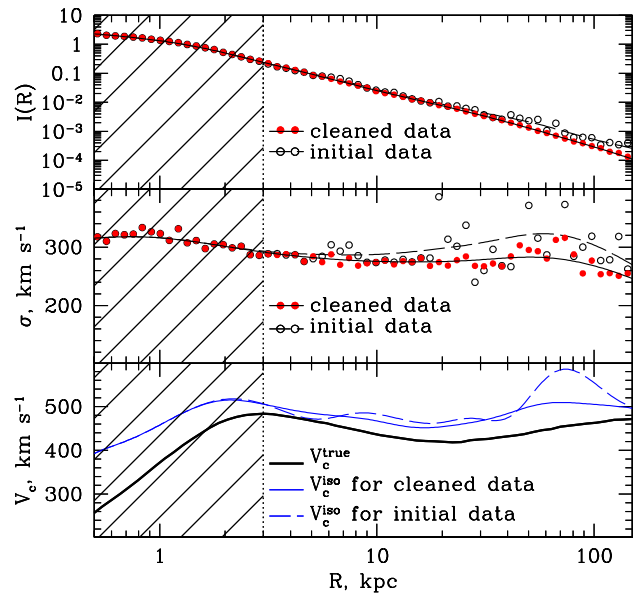}
\caption{Influence of satellites on the surface brightness (the upper panel) and the projected velocity dispersion profiles (in the middle). Open black circles correspond to the initial galaxy image and solid red circles - to the galaxy image without satellites. The black dashed curve is the smoothed curve for the initial data and the black solid curve is for the cleaned data. The bottom panel shows the true circular velocity (black thick line) and recovered circular velocity for the isotropic distribution of stellar orbits (in blue) for initial data (dashed) and cleaned data (solid). It is clear that removing satellites reduces the scatter in the line-of-sight velocity dispersion data and makes the profile smoother.      
\label{fig:M0053_opt}
}
\end{figure*}

Step 1: Excluding  satellites from the galaxy image.

Usually an image of a simulated galaxy (the distribution of stars projected onto a plane) contains many satellite objects and needs to be cleaned. Exclusion of satellites makes the surface brightness and the line-of-sight velocity dispersion profiles smoother and reduces the Poisson noise associated with satellites. The algorithm we used for removing satellites is as follows:
first, for each star a quantity $w$ characterising the local density of stars ($w \propto \rho_{*}^{-1/3}$) and analogous to the HSML (the SPH smoothing length) was calculated and the array of these values was sorted.  Then the $(0.4\cdot N_{stars})^{th}$ term of the sorted $w$-array was chosen as a reference value $w_o$. $N_{stars}$ is a total number of stars in a galaxy and a factor in front of $N_{stars}$ is some arbitrary parameter (the value $0.4$ was chosen by a trial-and-error method). Stars with the 3D-radius  $r > 10$ kpc and $w < w_o$ are considered as members of a satellite.  After projecting stars onto the plane perpendicular to the line of sight we have excluded all satellites together with an adjacent area of 1.5 kpc in size.
The inititial and final images of some arbitrarely chosen galaxy (the virial halo mass is $\simeq 1.7\times 10^{13} M_{\odot} h^{-1}$) are shown in Figure \ref{fig:M0053}.

Step 2: Evaluating $I(R)$ and $\sigma(R)$.

All radial profiles have been computed in a set of logarithmic concentric annuli around the halo center. 
To calculate the surface brightness profile, corrected for the contamination from the satellites, we have first counted the number of stars in each annulus, excising the regions around satellites. The surface area of each annuli has been also calculated, excluding the same regions. The ratio of there quantities gives us the desired `cleaned' surface brightness profile.  The average line-of-sight velocity of stars and the projected velocity dispersion have been calculated similarly. 

 Importance of the `cleaning' procedure and the resulting profiles of $I(R)$ and $\sigma(R)$ are shown in Figure \ref{fig:M0053_opt}. The surface brightness data (open circles correspond to the initial (`uncleaned') image and red solid circles  to the `cleaned' image) and the smoothed curves (the calculation of these curves is described in Step 3) are shown in the upper panel, the projected velocity dispersion profiles are shown in the middle panel. The true circular velocity $V_c^{\rm true}(r)$ (black solid curve) and recovered from the initial data (blue dashed line) and from `cleaned' data (blue solid line) circular velocity for the isotropic distribution of stellar orbits $V_c^{\rm iso}$ (the first equation in (\ref{eq:main})) are shown in the bottom panel. The last curve is in better agreement with the true velocity profile.       
All results and figures in this paper are restricted to the region $R > 3.0$ kpc.

Step 3: Taking derivatives.

To take derivatives  we follow the procedure described in \cite{2010MNRAS.404.1165C} in Appendix B. The main idea is that all data points participate in calculating the derivative but with different weights.
The weight function is given by 

\be
W(R_0,R)=\exp\left[-\frac{(\ln R_0-\ln R)^2}{2\Delta^2}\right], 
\label{eq:filter}
\ee 
where $R_0$ is the radius at which the derivative is being calculated and the parameter $\Delta$ is the width of the weight function. 

Both observed and simulated surface brightness profiles  are typically quite smooth so we have used $\Delta_I=0.3$ to calculate the logarithmic derivative $d\ln I(R)/ d\ln R$.  For the line-of-sight velocity dispersion data we have used $\Delta_{\sigma}=0.5$. With  the assumed values of $\Delta$ the local  perturbations are smoothed out but the global trend of the profiles is not affected. Changing values $\Delta_I$ and $\Delta_{\sigma}$ in the range $[0.3,0.5]$ does not significantly influence our  final result\footnote{If, however, we choose a width of the weight function smaller that $\Delta=0.3$ the local scatter in the data is not smoothed out and the results become ambiguous.}. The difference (in terms of circular velocity) is less than $1\%$. 
As an example the smooothed curves for the $I(R)$ and $\sigma(R)$ data in Figure \ref{fig:M0053_opt} are calculated using this procedure.

We have also tested the influence of parameters of the presented smoothing algorithm. As long as the smoothed curve describes data reasonably well neither the functional form of the weight function nor other parameters (like higher order terms in expansion $\ln I(R)=a(\ln R)^2+b\ln R +c$ or $\sigma (R)=a(\ln R)^2+b\ln R +c$) significantly affect the final result.

Step 4: Estimating the circular velocity.

Applying equations (\ref{eq:main})  or  (\ref{eq:agd_simple}) to the smoothed $I(R)$ and $\sigma(R)$  we have calculated $V_c$-profiles assuming isotropic, radial and circular orbits of stars. Then we have found a radius (a sweet point $R_{\rm sweet}$) at which the quantity $(V_c^{\rm iso}-\overline{V})^2+(V_c^{\rm rad}-\overline{V})^2+(V_c^{\rm circ}-\overline{V})^2$, where $\overline{V}=(V_c^{\rm iso}+V_c^{\rm rad}+V_c^{\rm circ})/3$, is minimal. The value of the isotropic velocity profile at this particular point is the estimation of the circular velocity speed we are looking for. We take $V_c^{\rm iso}$ as an estimate of the $V_c(R)$ (rather than $V_c^{\rm circ}$ or $V_c^{\rm rad}$) for two reasons. Firstly, at around one effective radius the dominant anisotropy for
most elliptical galaxies is $\disp \sigma_{zz} < \sigma_{RR} \sim \sigma_{\phi \phi}$ (\cite{2007MNRAS.379..418C}). The spherically averaged anisotropy is therefore only moderate (see also \cite{2001AJ....121.1936G}, Figure 4).  Massive elliptical galaxies are the most isotropic.  Thus an isotropic orbit distribution is a much better approximation than purely radial or circular orbits. Secondly, the value of $V_c^{\rm iso}$ is less prone to spurious wiggles in $I(R)$ and $\sigma(R)$.

\begin{figure}
\plotone{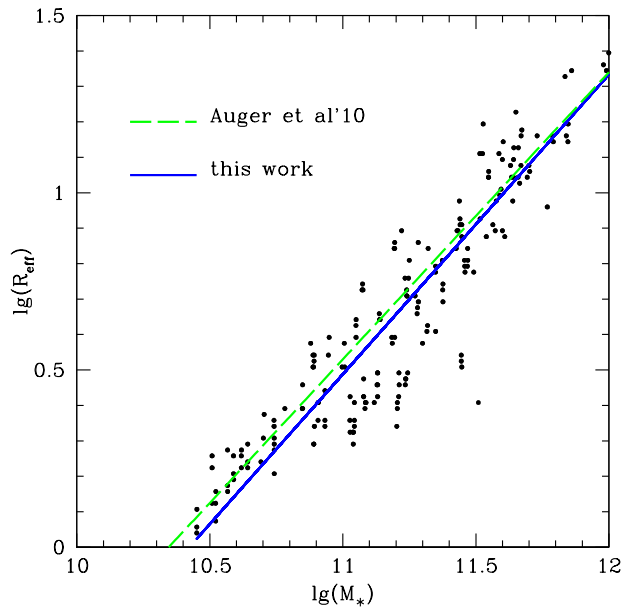}
\caption{$R_{\rm eff} - M_{\ast}$ relation. The blue solid line is the linear fit to data points from the simulations. The green dashed line is the observed mass-size relation from \citep{2010ApJ...724..511A}.
\label{fig:reff}
}
\end{figure}

The effective radius $R_{\rm eff}$ is calculated as a radius of the circle which contains half of the projected stellar mass, taking into account effects of cleaning.  We found that in the simulated data-set the value of the effective radius depends on the maximal radius used to calculate the total number of stars in a galaxy. The problem is  especially severe for the most
 massive galaxies as they have an almost power-law 3D stellar density distribution $\disp \rho_{\ast}\propto r^{-a}$ with $a \simeq 3$. In our analysis (in contrast to \cite{2011arXiv1106.5490O}) we have not introduced any artificial cut-off and used all stars  in the smooth stellar component (excluding substructure) of the main galaxies out to their virial radii for the calculation of the effective radius. The resulting effective radii as a function of total stellar mass (in logarithmic scale) are shown in Figure (\ref{fig:reff}). The slope and the normalization of the $R_{\rm eff}-M_{\ast}$ relation are close to the fit of SLACS data by \cite{2010ApJ...724..511A}. 

The axis ratio $q$ of each projection of a galaxy is calculated as a square root of eigenvalues of the diagonalised inertia tensor.  The inertia tensor is computed within the effective radius without excluding substructures. We have found that $q$ is not sensitive to our cleaning procedure as normally there are almost no satellites within $R_{\rm eff}$.

\section{Analysis of the sample} 
\label{sec:analysis}

\subsection{At a sweet point} 
\label{subsec:points}

\begin{figure*}
\plotone{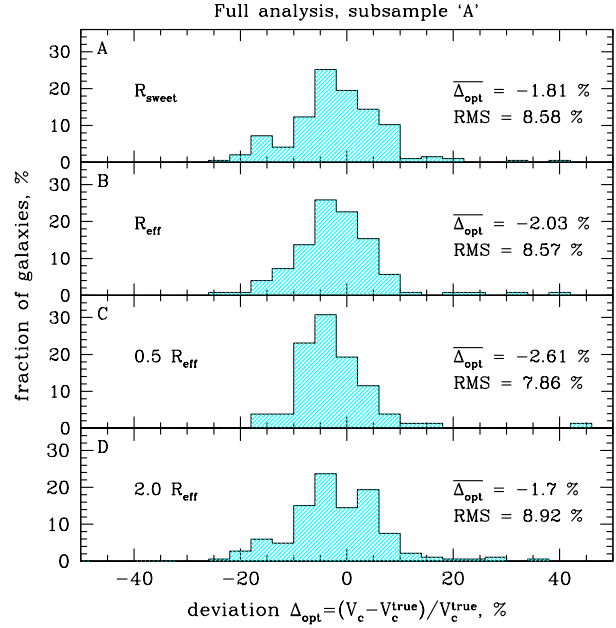}
\caption{The fraction of galaxies (in \%) as a function of deviation $\Delta_{opt}=\left(V_c^{\rm iso}-V_c^{\rm true}\right)/V_c^{\rm true}$ evaluated via equations (\ref{eq:main}) at different radii: $R_{\rm sweet}$ (panel (A)), $R_{\rm eff}$ (panel (B)), $0.5 R_{\rm eff}$ (panel (C)) and $2R_{\rm eff}$ (panel (D)). 
\label{fig:all}
}
\end{figure*}
\begin{figure*}
\plottwo{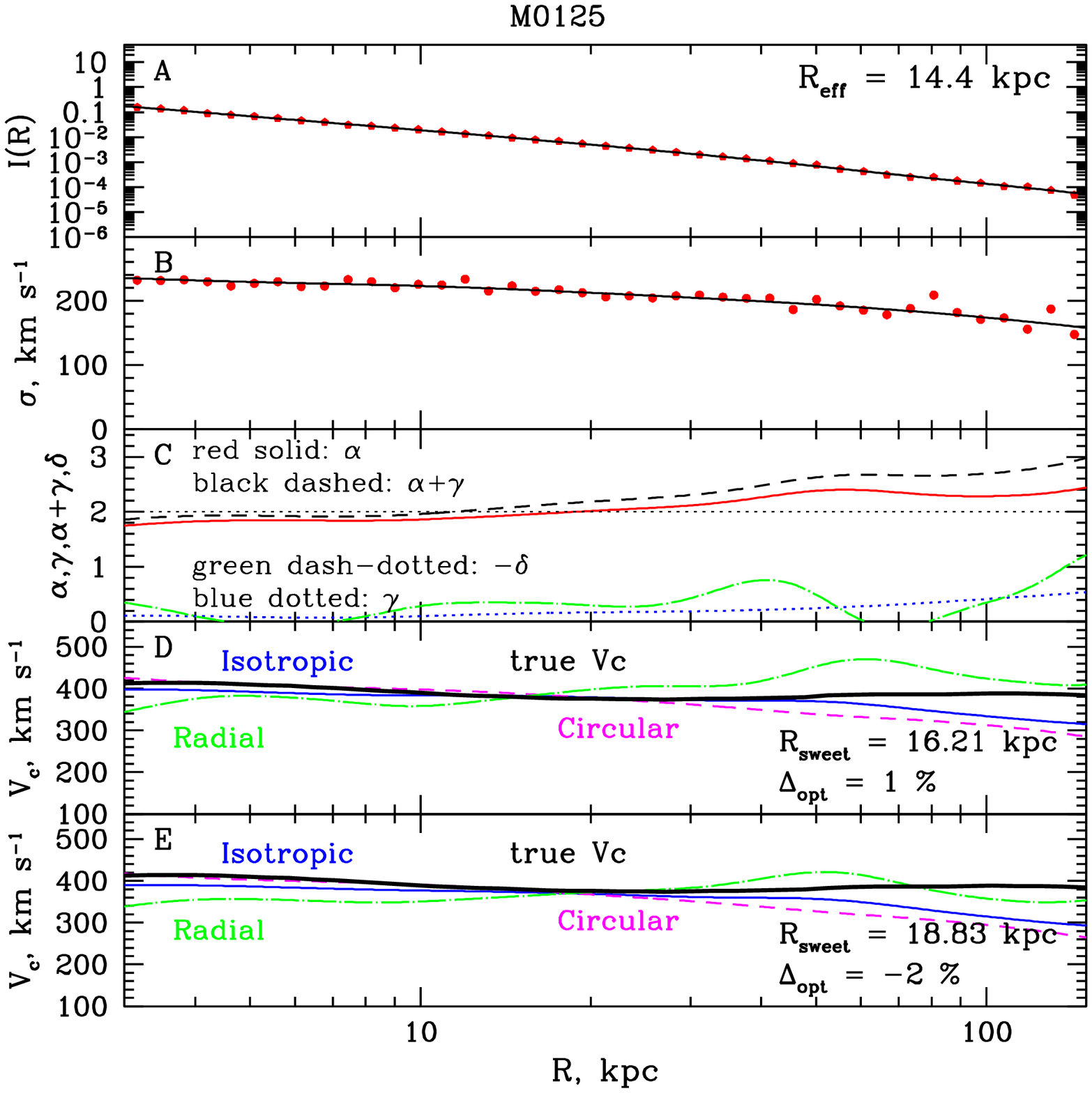}{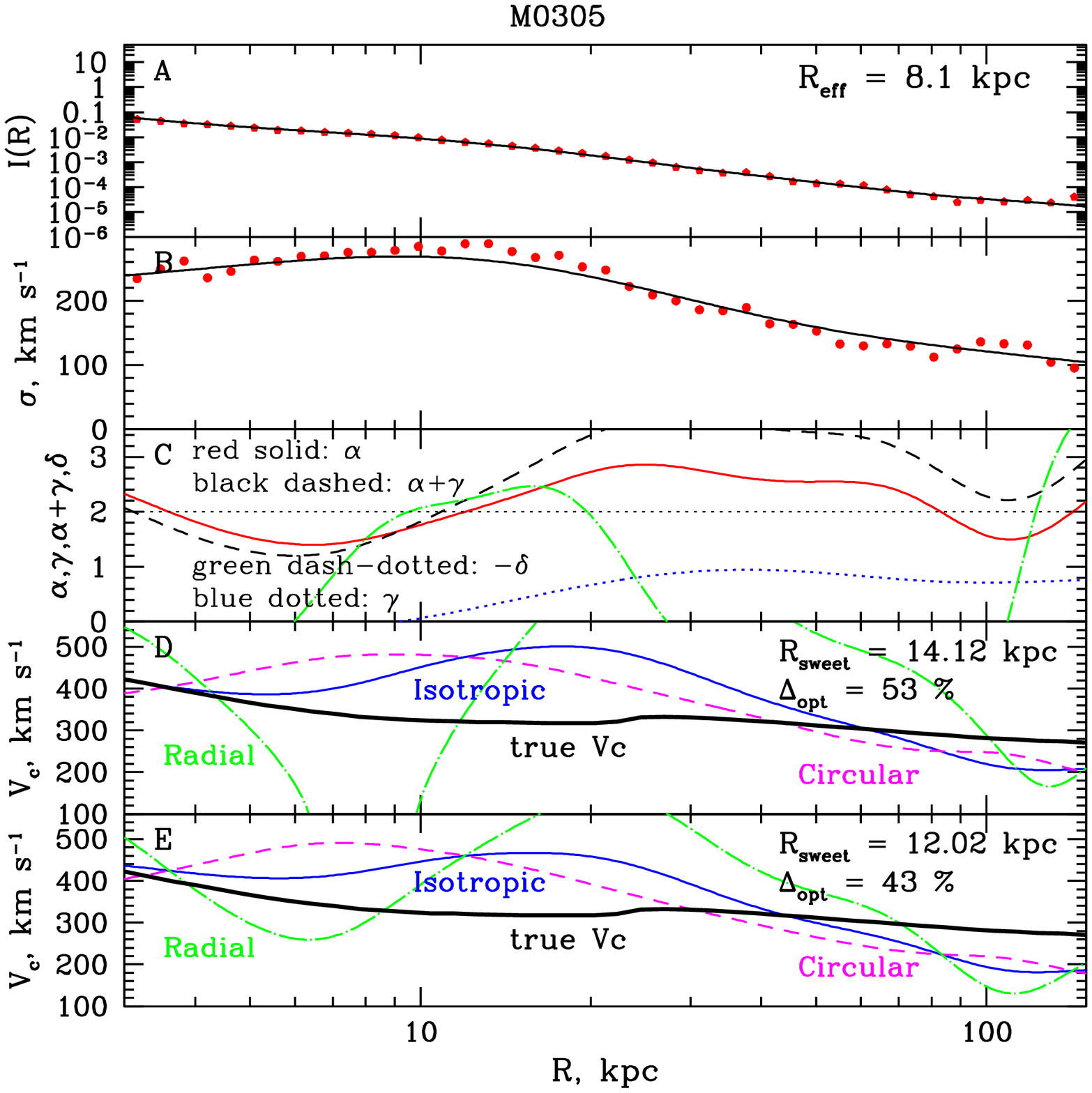}
\caption{{Left:} Example of the galaxy that perfectly suits for the analysis. The surface brightness and the projected velocity dispersion profiles are shown in panels (A) and (B) correspondingly. Data are represented as red points and smoothed curves that were used to compute derivatives ($\alpha, \gamma, \delta$) as black solid lines. The auxilary coefficients $\alpha, \gamma, -\delta$ and $\alpha+\gamma$ are shown in panel (C) in red solid, blue dotted, green dash-dotted and black dashed lines, respectively. Circular velocity profiles for isotropic orbits of stars (blue solid line), pure radial (green dash-dotted) and pure circular (magenta dashed) orbits as well as the true circular speed (black thick curve) are presented in panel (D) for the full version of the analysis (equations (\ref{eq:main})). And the same curves for the simplified analysis (equations (\ref{eq:agd_simple})) are shown in panel (E). 
{Right:} Example of the galaxy with large deviation $\Delta_{opt}$ due to merger activity. The crest in the projected velocity dispersion profile at $R \simeq 20$ kpc leads to the significatly overestimated value of the circular speed.
\label{fig:exmerg}
}
\end{figure*}
\begin{figure*}
\plottwo{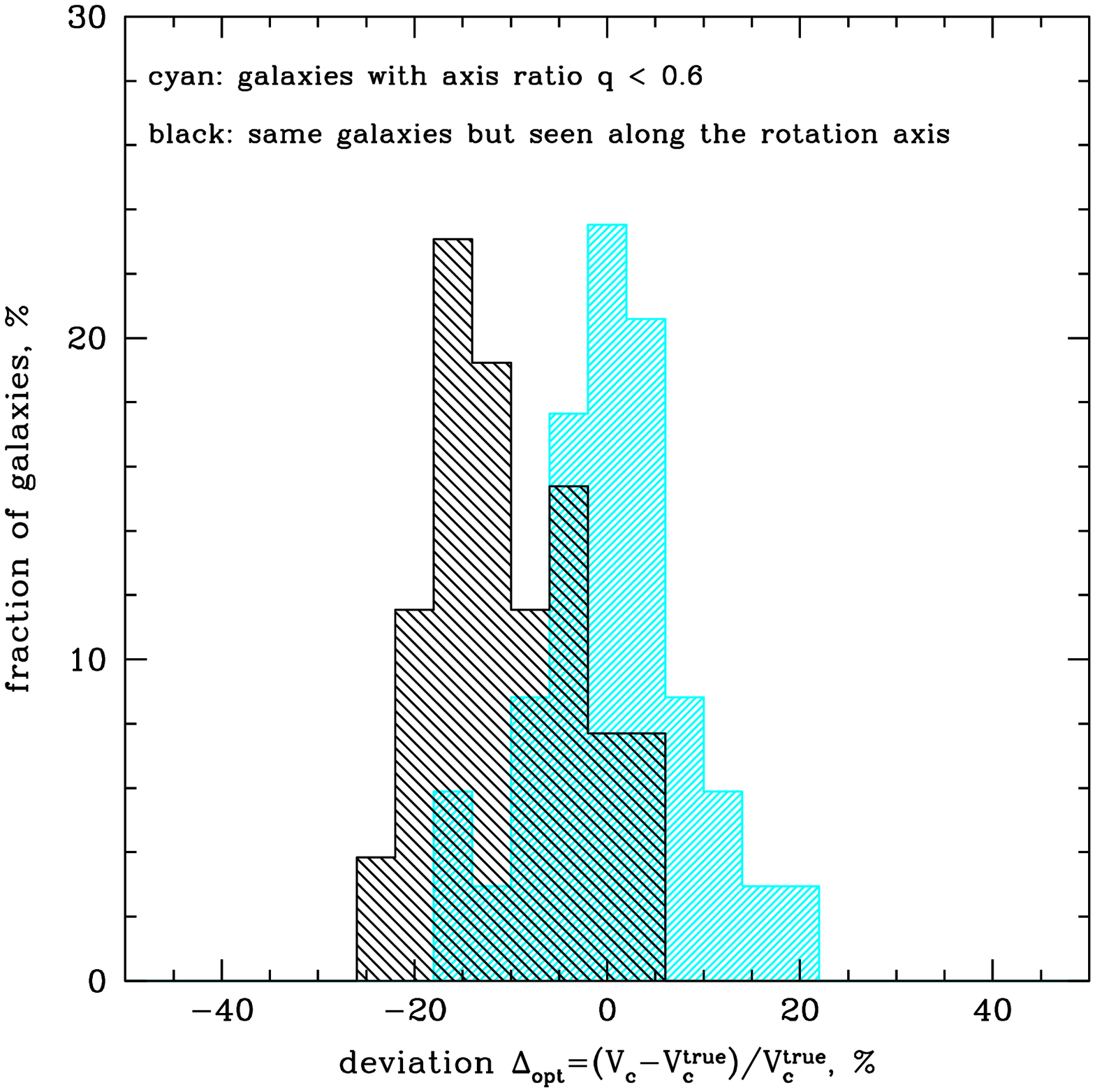}{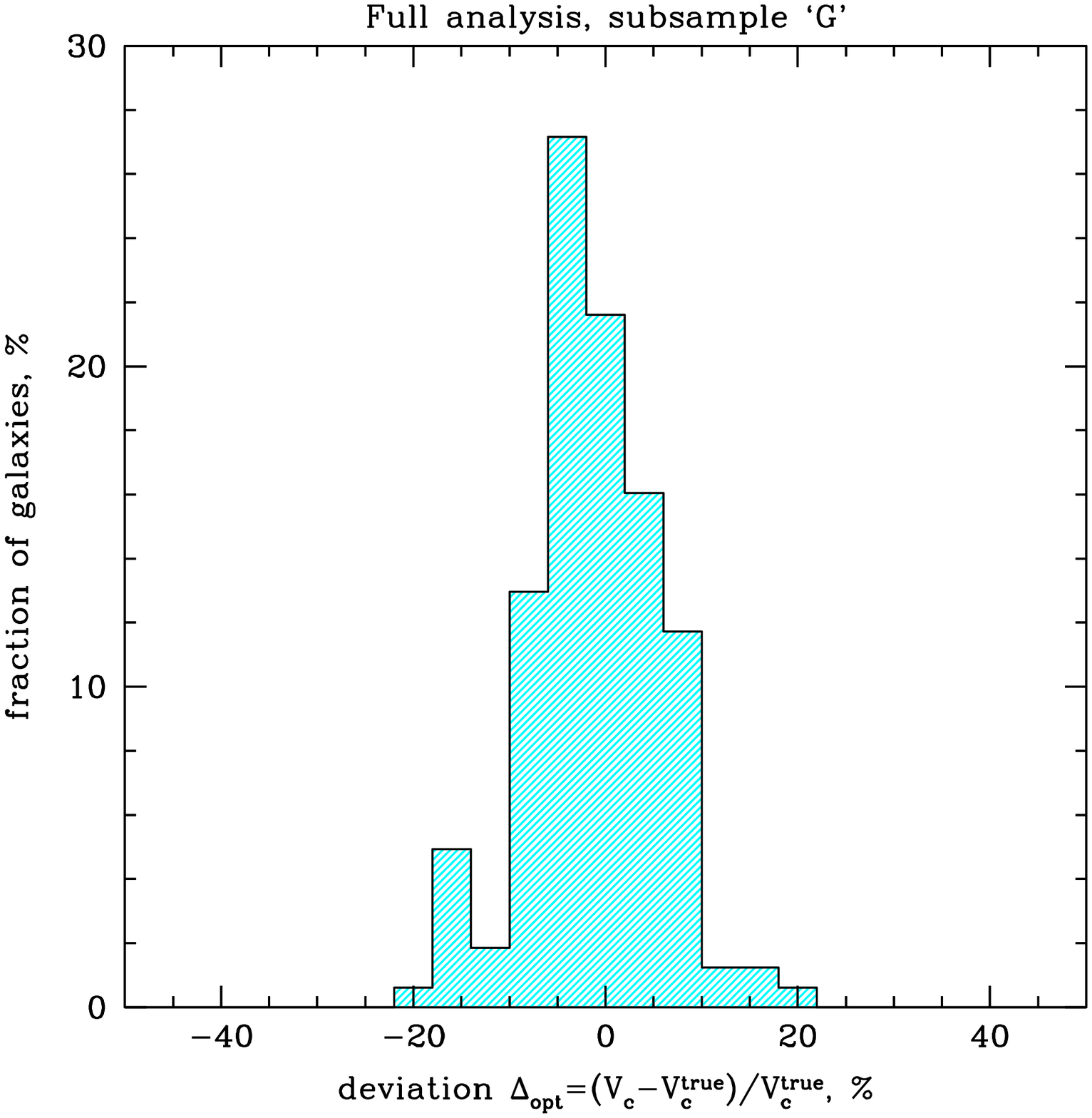}
\caption{{Left: Shown in cyan is the histogram for deviations for galaxies with the axis ratio $q < 0.6$, in black is the histogram for the same galaxies but seen in a projection with the axis ratio $q$ close to unity (= seen along the rotation axis).}
{Right: The histogram for deviations for the sample when merging and oblate galaxies seen along the rotation axis are excluded (subsample `G'). The average deviation $\overline{\Delta_{opt}}=(-1.2 \pm 0.9) \%$, $RMS = 6.8 \%$. }
\label{fig:triax}
}
\end{figure*}

For each galaxy in the sample we have performed all steps described above and we have selected the radius at which the circular velocity curves for isotropic, circular and radial orbits (equations (\ref{eq:main})) intersect or lie  close to each other. Then we have calculated the value of the isotropic speed $V_c^{\rm iso}$ at this radius. To measure the accuracy of our estimates let us introduce a deviation from the true circular speed $\Delta_{opt}=\left(V_c^{\rm iso}-V_c^{\rm true}\right)/V_c^{\rm true}$, where $V_c^{\rm iso}$ and $V_c^{\rm true}$ should be taken at the sweet spot $R_{\rm sweet}$. The subscript `opt' (=~optical) is used to distinguish this method (based on optical data) from circular speed calculations based on X-ray data.
We have plotted the number of galaxies (normalised to the total number of galaxies and expressed in $\%$) versus the deviation $\Delta_{opt}$ in a form of a histogram. To have an idea whether the method under consideration gives resonable accuracy, histograms for deviations at $R_{\rm eff}$, $0.5 R_{\rm eff}$ and $2R_{\rm eff}$ are also shown. The whole sample (`subsample A') is presented in Figure \ref{fig:all}. The sample averaged value of the deviation $\overline{\Delta_{opt}}$ is slightly less than zero in all cases. For example, at the sweet point $\overline{\Delta_{opt}}=(-1.8 \pm 1.1) \%$ while the RMS $= 8.6 \%$\footnote{$\overline{x}=\disp \frac{\sum{x}}{N}$, $RMS=\sqrt{\disp \frac{\sum{(x-\overline{x})^2}}{N-1}}$}.

Large deviations ($\sim 30-40\%$) are seen only in galaxies with ongoing merger activity. The influence of mergers appears as `waves' in the projected velocity dispersion profile. The example of such a system is shown in Figure \ref{fig:exmerg} (right panel). The presence of such `waves' indicates that the circular speed  could be significantly overestimated (by a factor of $\sim 1.2-1.5$), which is not surprising as the method is based on the spherical Jeans equations and the assumption about dynamical equilibrium is violated.  When the profiles $I(R)$ and $\sigma(R)$ are smooth and monotonic the circular speed can be recovered with much higher accuracy (Figure \ref{fig:exmerg}, left panel).

The sample includes galaxies with different values of ellipticity. The axis ratio $q$ (computed from the diagonalized inertia tensor within $R_{\rm eff}$) ranges from $0.19$ to $0.99$. To test the possible influence of the ellipticity on the accuracy of estimates we have selected galaxies with axis ratio $q < 0.6$. The resulting distribution as a function of the circular speed deviations is almost symmetric, unbiased, with $RMS \simeq 8\%$ (Figure \ref{fig:triax}). On the other hand, if we consider the same galaxies seen in a projection with the maximum value of the axis ratio $q$, we get the distribution appreciably biased toward negative values of the deviation ($\overline{\Delta_{opt}}=(-10.2 \pm 1.6) \%$). The reason for this bias is rotation. When observing a galaxy along its rotation axis the projected velocity dispersion is appreciably smaller than for perpendicular directions. To further test this statement we have rotated each galaxy so that the principal axes of the galaxy ($A\ge B \ge C$) coincide with the coordinate system ($x$, $y$ and $z$, correspondingly) and analysed velocity maps for each projection. As a criteria for rotation we have used the anisotropy-parameter $(v/\sigma)^{*}=\disp \frac{v/\overline{\sigma}}{\sqrt{(1-q)/q}}$, where $v$ is the average rotation velocity of stars, $\overline{\sigma}$ is the mean velocity dispersion and $q$ is the axis ratio \citep{1978MNRAS.183..501B, 1990A&A...239...97B}. If $(v/\sigma)^{*}>1.0$ then the object is assumed to be rotating. We have found that the most massive simulated galaxies usually do not rotate or rotate slowly and show signs of triaxiality while less massive galaxies rotate faster and show signs of axisymmetry. This statement is in agreement with observational studies 
(e.g. \cite{2007MNRAS.379..418C} and references therein). Moreover, the majority of rotating galaxies appears to be oblate, rotating around the short axis. So for the oblate galaxies observed along the rotation axis (and as a consequence seen in a projection with the axis ratio $q$ close to unity) the method gives underestimated values of the circular speed. It should be noted that when observing the rotating galaxies along long axes the circular speed estimate is slightly biased towards overestimation (\cite{2007MNRAS.381.1672T} reached the similar conlusion).  The average deviation for the subsample of oblate galaxies seen perpendicular to the rotation axis is biased high by $\Delta_{opt}\simeq (4.4 \pm 1.4) \%$ with RMS $=6.3 \%$.

To investigate possible projection effects on the results of our analysis we have picked one rotating galaxy (the virial halo mass is $\simeq 2.2\times 10^{12} M_{\odot} h^{-1}$) and calculated the surface brightness and the velocity dispersion profiles for different lines of sight. While the light profiles are quite similar, the velocity dispersion profiles may differ significanly when the line of sight is parallel to the rotation axis and perpendicular to it. We have calculated the average value of the circular speed estimates taking into account the probability of observing the galaxy at different angles. For the selected galaxy the average deviation from the true $V_c$ is about $-4.9 \%$ and the maximum deviation (when observing along the rotation axis) is about $-25 \%$.

\begin{figure*}
\plottwo{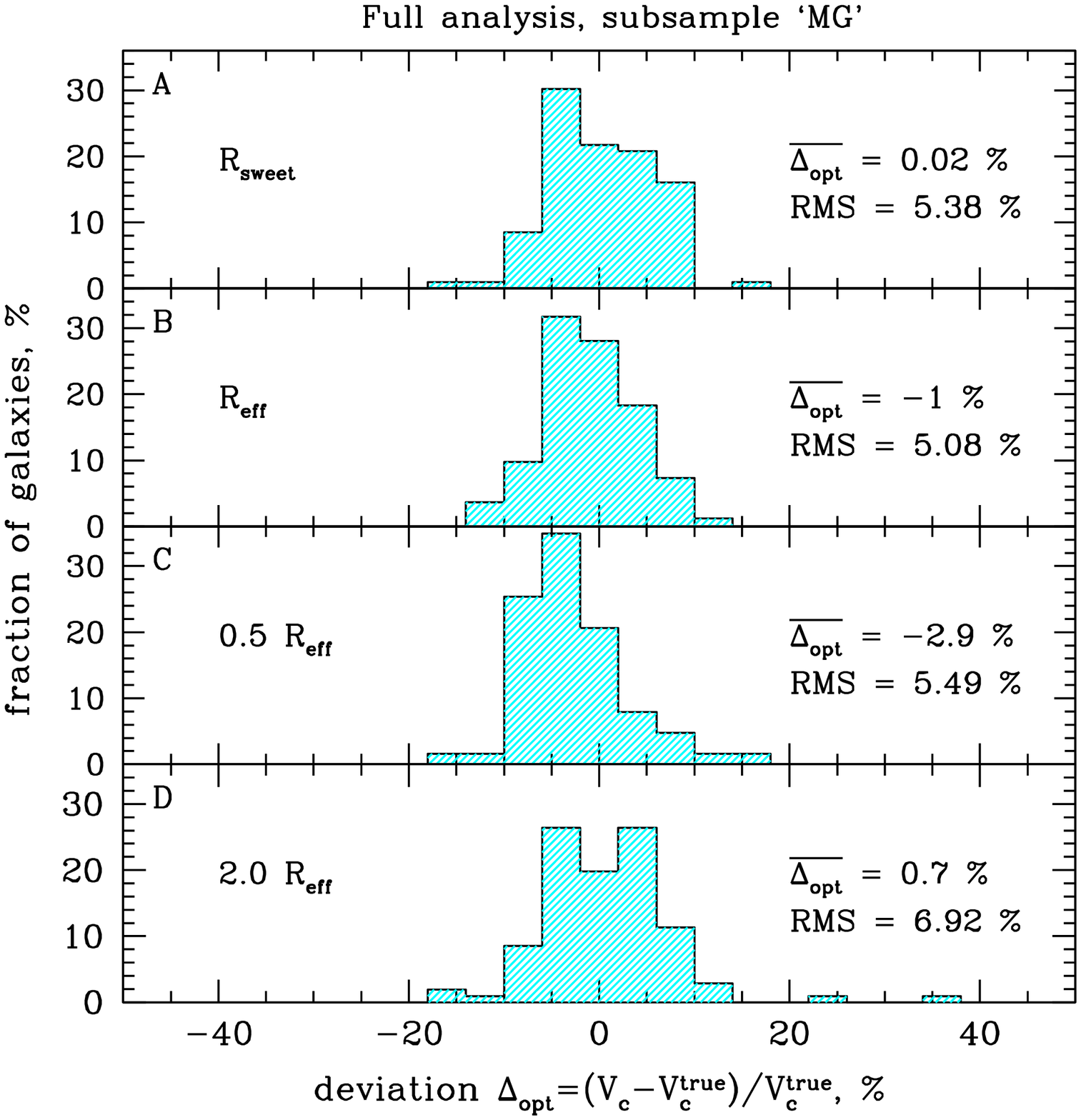}{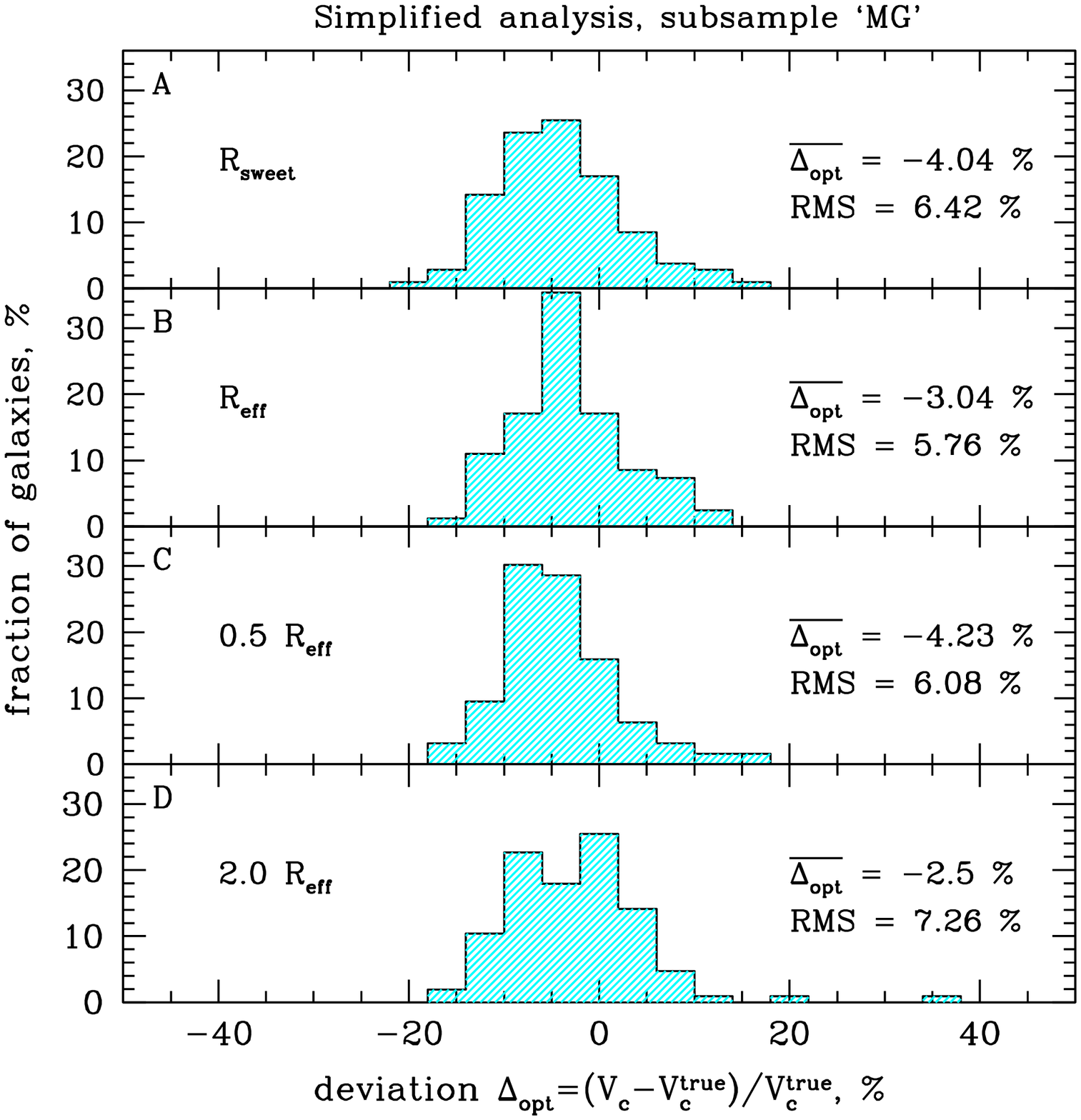}
\caption{{Left:} Distribution of galaxies from the subsample `MG' (massive galaxies with $\sigma(R_{\rm eff}) > 150$ $\kms$ when merging and oblate galaxies observed along the rotation axis are excluded) according to their deviations. Deviations are calculated at $R_{\rm sweet}$ (panel (A)), $R_{\rm eff}$ (panel (B)), $0.5R_{\rm eff}$ (panel (C)) and $2R_{\rm eff}$ (panel (D)). 
{Right:} The same histograms but for the simplified version of the analysis (equations (\ref{eq:agd_simple}))
\label{fig:MG}
}
\end{figure*}

It should be mentioned that the method under consideration was designed for recovering the circular speed in massive elliptical galaxies and it does not pretend to give accurate results for low-mass galaxies. In addition, not so many elliptical galaxies with $\sigma < 150-200 \kms$ are observed \citep[e.g.][]{2010MNRAS.404.2087B}.

It is convenient to distinguish low and high mass simulated galaxies by the value of the projected velocity dispersion at the effective radii. Let us call `massive' galaxies with $\sigma(R_{\rm eff})>150 \kms$.  If we apply our analysis to the subsample of massive galaxies and exclude merging and oblate galaxies seen along the rotation axis (the subsample `MG'), we get an unbiased distribution with $RMS = 5.4 \%$. The resulting histogram is shown in Figure \ref{fig:MG}, left image, panel (A). Estimations at other radii give slightly more biased and slightly less accurate results (Figure \ref{fig:MG}, left image, panels (B)-(D)).  

Thereby we have marked out four subsamples - the whole sample without exceptions (`A' - all), the sample without merging or oblate galaxies seen along the rotation axis (`G' - good), the subsample of massive galaxies (`M' - massive) with $\sigma(R_{\rm eff})>150$ $\kms$ and, finally, the subsample of massive galaxies when merging and oblate galaxies observed along the rotation axis are excluded (`MG' - massive and good).

In case of missing or unreliable data on the line-of-sight velocity dispersion profile \cite{2010MNRAS.404.1165C} suggest to apply a simplified version of the aforementioned analysis (equations (\ref{eq:agd_simple})). By neglecting terms $\gamma$ and $\delta$ we assume that the projected velocity dispersion profile is flat. Then the radius at which $I(R)\propto R^{-2}$ is the sweet point. The resulting histograms for the subsample `MG' are shown in Figure \ref{fig:MG}, right panel. It can be seen that data on the projected velocity dispersion plays noticable role in the analysis if the required accuracy is of order of several $\%$. Neglecting its
 derivatives leads to a bias towards underestimated values of $V_c$ ($\overline{\Delta_{opt}}=(-4.0 \pm 1.1) \%$ at the sweet point) and broader wings/tails (RMS = $6.4 \%$ at $R_{\rm sweet}$) compared to Figure \ref{fig:MG}, left panel. Nonetheless, if only the surface brightness profile and some data on the projected velocity dispersion are available the simplified version of the method seems to be a good choice.

\subsection{Simulated galaxies at high redshifts}
\label{subsec:highred}

We have also tested the same procedure for galaxies at higher redshifts. Namely, at $z=1$ and $z=2$. The fraction of merging galaxies in the sample is larger at high redshift than at $z=0$ and the number of stars in each halo is considerably smaller. Nevertheless, results are quite encouraging. For the subsample `MG' the average deviation of the circular speed for the isotropic distribution of orbits at the sweet point (estimated via equations (\ref{eq:main})) from the true one is close to zero and the scatter is modest. At redshift $z=1$ the average deviation is $\overline{\Delta_{opt}}=(-0.3\pm 1.1) \%$ and RMS = 6.0 \%, at $z=2$ $\overline{\Delta_{opt}}=(0.9 \pm 2.2)\%$ and RMS = 8.0 \%.

\subsection{Mass from integrated properties}
\label{subsec:potential}

\begin{figure*}
\plottwo{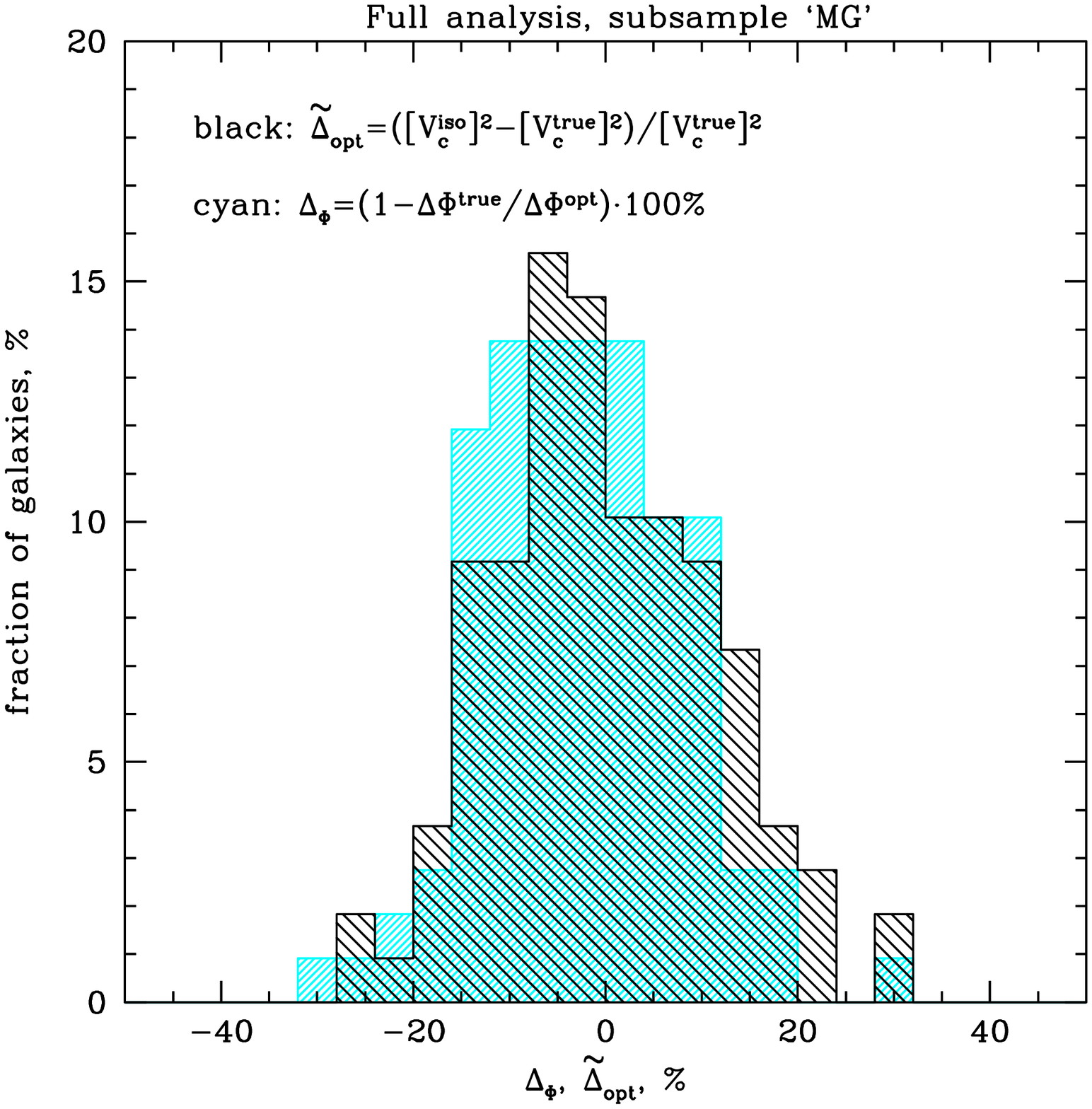}{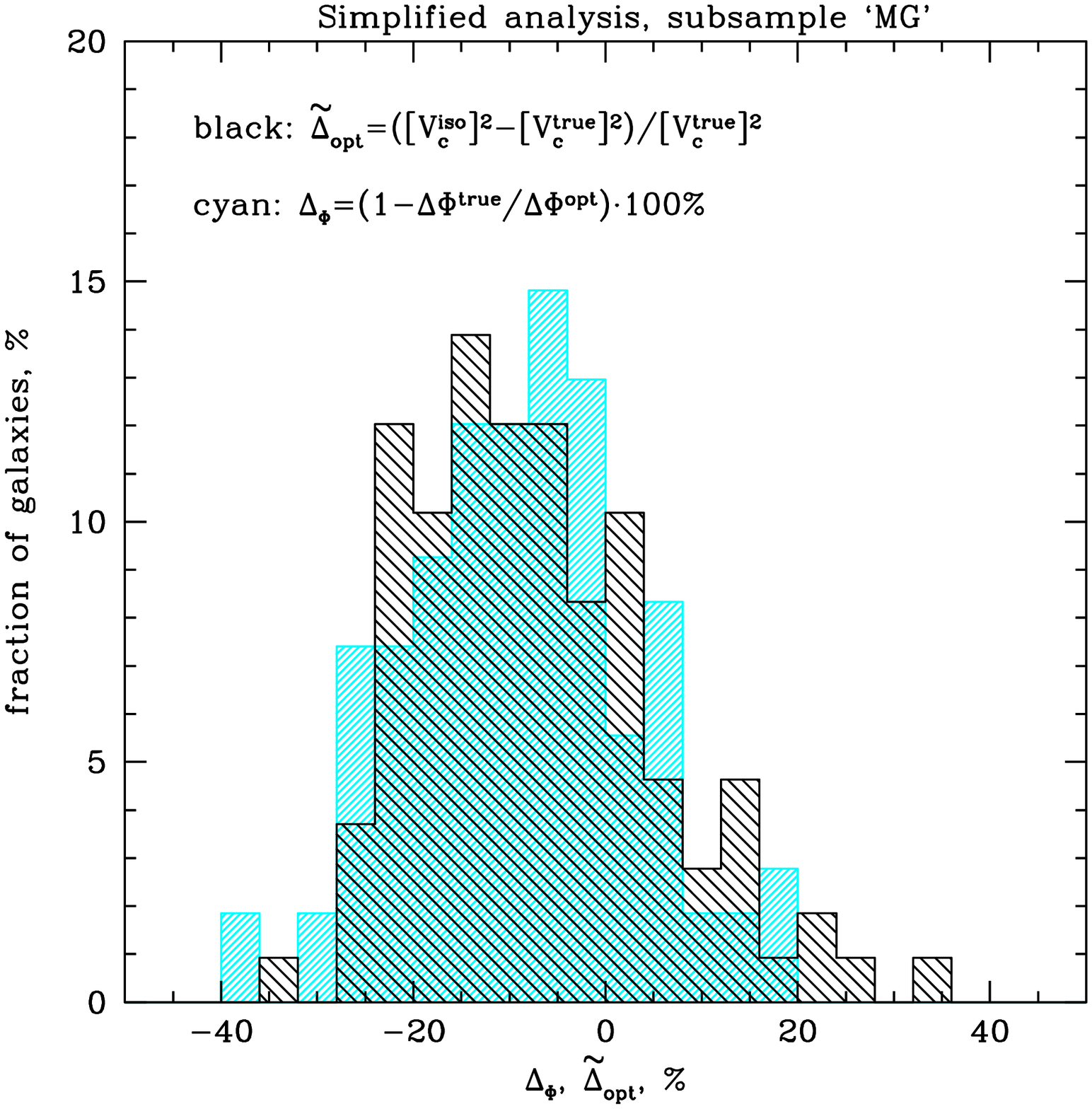}
\caption{Accuracy of the derived potential of massive galaxies (merging and oblate objects seen along the rotation axis are excluded). In cyan shown the histogram for the quantity $\Delta_{\Phi}=(1-\kappa)\cdot 100 \%$, where $\kappa = \disp \Delta \Phi^{true}/\Delta \Phi^{opt}$. In black shown the histogram for the deviation $\tilde{\Delta}_{opt}$ of the estimated at the sweet point $\left[V_c^{\rm iso}\right]^2$ from the true one $\left[V_c^{\rm true}\right]^2$. {Left:} Histograms for the full version of the analysis (equations (\ref{eq:main})). The average value of $\kappa$ is $\overline{\kappa} = 1.02 \pm 0.02$ and $RMS = 10.3 \%$. The average value of the deviation $\tilde{\Delta}_{opt}$ is $(-0.2 \pm 1.9) \%$ and RMS = $11.3 \%$. {Right:} Histograms for the simplified version of the analysis (equations (\ref{eq:agd_simple})). $\overline{\kappa} = 1.09 \pm 0.02$ and RMS $=11.8 \%$, $\overline{\tilde{\Delta}_{opt}}=(-7.2 \pm 2.1) \%$ and RMS = $12.7 \%$.
\label{fig:potmass}
}
\end{figure*}

Asssuming the logarithmic form of the gravitational potential $\Phi(r)=V_c^2\ln r+const$ we can estimate the potential $\Phi$ over some range of radii up to a constant.
To calculate the potential one needs to know the circular velocity profile. If we assume that this profile roughly coincides with the isotropic profile $V_c^{\rm iso}$ over some range of radii (let us choose $R \in [0.5R_{\rm eff}, 3R_{\rm eff}]$ as a range of radii easily available for observations), we can define:  
\be
\Phi^{opt} = \disp\int\limits_{0.5R_{\rm eff}}^{R} \frac{\left[V_c^{\rm iso}\right]^2}{r}\,dr,
\label{eq:pot}
\ee 
where $R \in (0.5R_{\rm eff},3R_{\rm eff})$ and $V_c^{\rm iso}$ can be found from the full version of the analysis (the first equation of (\ref{eq:main})) or from the simplified version (the first equation of (\ref{eq:agd_simple})). As the true potential is known we can write $\Phi^{\rm true}=\kappa \cdot \Phi^{opt}+{\rm const}$. In the ideal case  $\kappa = \disp \Delta \Phi^{true} / \Delta \Phi^{opt} = 1.0$. The accuracy of such an approach is illustrated in Figure \ref{fig:potmass}. In cyan is shown the distribution of subsample `MG' of galaxies as a function of $\Delta_{\Phi}=(1-\kappa)\cdot 100\%$. Just to remind this subsample consists of the massive galaxies with $\sigma(R_{\rm eff})>150$ $\kms$ and merging galaxies as well as oblate galaxies seen along their rotation axes are excluded. In case of the full analysis the distribution is almost unbiased (the average value of $\kappa$ is $1.02 \pm 0.02$) with $RMS = 10.9 \%$. For the simplified formula of $V_c^{\rm iso}$ we see some offset $\overline{\kappa} = 1.09 \pm 0.02$ and  RMS = $11.8 \%$. 
In approximation of small deviations RMS defined for the potential calculations is twice as large as RMS for the circular velocity calculations because the potential $\Phi$ scales as  $ V_c^2$.
To compare this approach with previous results let us introduce the deviation $\tilde{\Delta}_{opt}=(\left[V_c^{\rm iso}\right]^2-\left[V_c^{\rm true}\right]^2)/\left[V_c^{\rm true}\right]^2$ estimated at the sweet point $R_{\rm sweet}$. Resulting distribution for the same subsample is shown in black in Figure \ref{fig:potmass}. As expected the width of this distribution is nearly two times larger than the distribution of circular velocity estimates (Figure \ref{fig:MG}).

As we see the gravitational potential can be estimated via $V_c^{\rm iso}$ with reasonable accuracy. This fact is in agreement with the aforementioned statement that most massive galaxies in the sample have almost flat circular velocity profiles in broad range of radii.

\subsection{Circular speed derived from the projected dispersion in a fixed aperture}
\label{subsec:aperture}

\begin{figure*}
\plotone{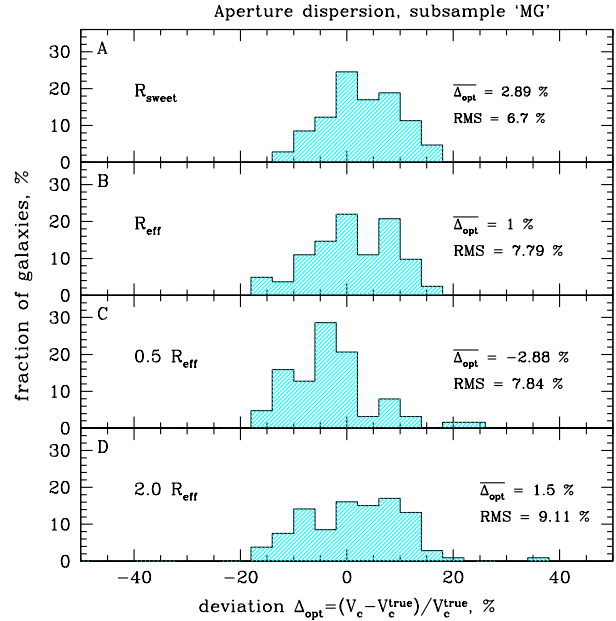}
\caption{Distribution of galaxies from the subsample `MG' (galaxies with $\sigma(R_{\rm eff}) > 150$ $\kms$ when merging and oblate galaxies seen along the rotation axis are excluded).
\label{fig:aperture}
}
\end{figure*}

When data on the velocity dispersion are available only in the form of aperture dispersions one can estimate the circular speed using the simple relation

\be
V_c^2=3 \cdot \sigma_{ap}^2(<R),
\label{eq:vcap}
\ee

where $\sigma_{ap}(<R)$ is the velocity dispersion measured within some aperture $R$. To test this way of evaluating the circular speed we have computed the luminosity-weighted dispersion within the aperture of radius $R$ (excluding the region $R<2R_{soft}=3$ kpc where softetning may affect results of the analysis)
\be
\sigma_{ap}^2(<R)=\frac{\int\limits_{2R_{soft}}^{R} I(x)\sigma^2 (x) xdx}{\int\limits_{2R_{soft}}^{R} I(x)xdx}
\label{eq:aperture}
\ee 

and calculated the deviation from the true circular speed at different radii: at the sweet point $R_{\rm sweet}$  for the full version of the analysis (equations (\ref{eq:main})), at $R=0.5R_{\rm eff}, R_{\rm eff}$ and $2R_{\rm eff}$. The resulting histograms for the subsample `MG' are presented in Figure \ref{fig:aperture}. Comparing with circular speed estimations at a single radius (in particular, at the sweet point) this method gives a biased result $\overline{\Delta_{opt}}(R_{\rm eff})=(1.0 \pm 1.3) \%$ and noticebly larger RMS (at $R_{\rm eff}$ RMS = 7.8\%).

\subsection{Circular speed from X-ray data}
\label{subsec:sxray}

Another way to measure the mass of galaxies comes from the analysis of
the hot X-ray gas in galaxies. By measuring the gas number density $n$ and
the temperature $T$ profiles from X-ray observations one can find the total
mass assuming that the gas is in the hydrostatic equilibrium (HE) \citep[e.g.][]{Mat78,For85}. Assuming spherical symmetry, the equation of 
HE can be written
\be
-\frac{1}{\rho}\frac{dP}{dr}=\frac{d\Phi}{dr}=\frac{V_c^2}{r}=\frac{GM}{r^2},
\label{eq:he}
\ee 
where $P=nkT$ is the gas pressure ($n$ is the gas number density), $\rho=\mu m_p n$ is the gas
density ($m_p$ is the proton mass), $\Phi$ is the gravitational potential, $V_c$ is a circular
velocity and $M$ is the total mass of the galaxy.  In simulations the mean atomic weight $\mu$ is assumed to be equal to 0.58. Strictly speaking,
assuming the HE one neglects possible non-thermal contribution to the
pressure, which can be due to presence of (i) turbulence in the
thermal gas, (ii) cosmic rays and magnetic fields \citep[e.g.][]{2008MNRAS.388.1062C}.
     
To estimate deviations from HE, the so called mass bias, we
took a subsample of the most massive galaxies with $M > 6.5\cdot 10^{12} M_\odot$. X-ray properties of low mass galaxies in the sample are influenced by gravitational softening in the central
3-4 kpc and  are strongly dominated by cold and
dense clumps in center. Moreover, we know from observations that only the most
massive galaxies have massive X-ray atmospheres \citep[e.g.][]{2001MNRAS.328..461O}.

\begin{figure*}
\plotwide{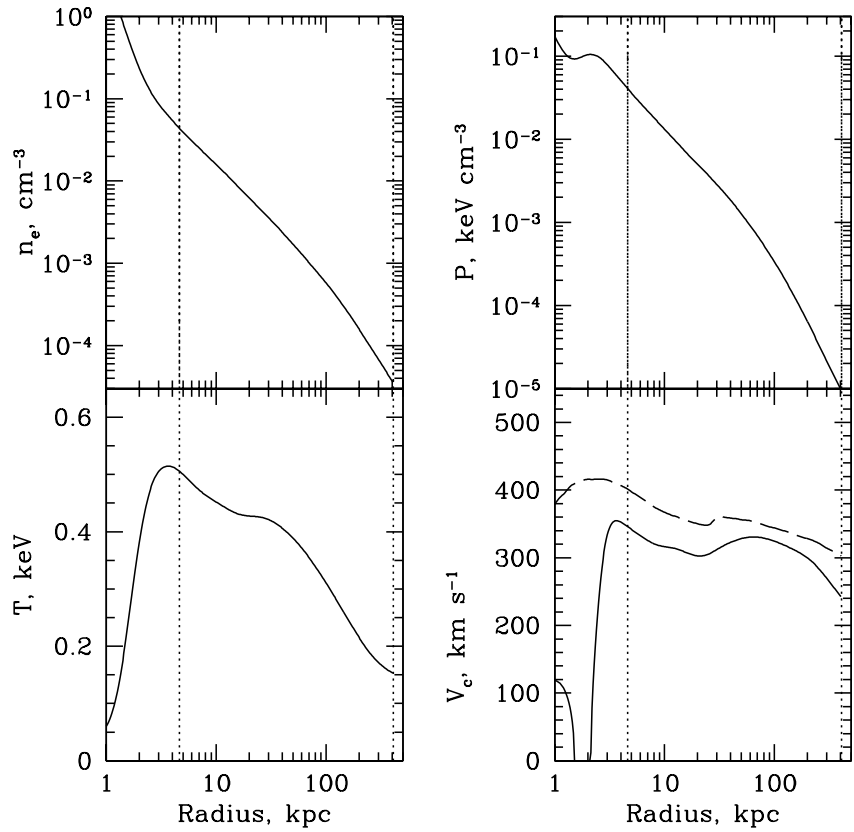}
\caption{ Profiles of hot gas electron number density, temperature, pressure
  and circular velocity of simulated galaxy. Dotted vertical curves
  show the upper and lower limits on $R$. $V_c$ plot: solid and
  dashed curves show mass from HE and total mass from simulations respectively. 
\label{fig:xprof}
}
\end{figure*}

The typical profiles of the gas density and temperature extracted from
simulations are shown in Figure \ref{fig:xprof}. We used the median
value of the electron density $n_e$ and $T$ determined in each spherical shell, so that we are free of
cold dense clumps contamination \citep{Zhu11}. Calculated pressure and circular
velocity (equation \ref{eq:he}) are also shown in Figure \ref{fig:xprof}.  The
spurious feature of simulations is that in the central 3-4 kpc cold
and dense clumps are strongly dominating. Even using the median value
does not remove these clumps, causing strong increase of density and
drop of temperature in the center.  These clumps are moving ballistically
and are not in the HE. 

\begin{figure}
\plotone{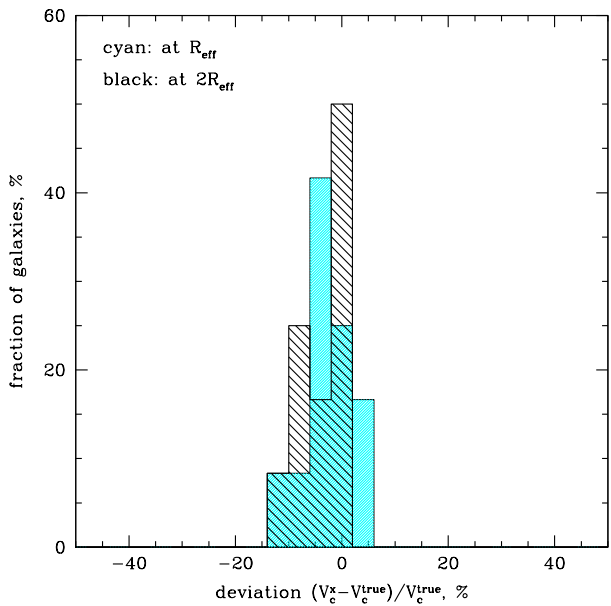}
\caption{Distribution of galaxies according to their deviations of estimated circular speed from the true value at $R_{\rm eff}$ (shown in cyan) and at $2R_{\rm eff}$ (shown in black). Circular speed is derived using the hydrostatic equilibrium equation for the hot gas. The sample consists of $12$ galaxies.
\label{fig:hyd}
}
\end{figure}

Deviations from HE $\disp\Delta_X=\frac{V_c-V_c^{\rm true}}{V_c^{\rm
    true}}$ were calculated at $R_{\rm eff}$ (cyan histogram in
Figure \ref{fig:hyd}) and $2R_{\rm eff}$ (black histogram in
Figure \ref{fig:hyd}). The average over the subsample value of the deviation at $R_{\rm
  eff}$ is $\disp \overline{\Delta_X}=(-3.0 \pm 1.3) \%$ and RMS=$4.4 \%$. At 2$R_{\rm eff}$ $\disp \overline{\Delta_X}= (-4.0 \pm 1.1) \%$, RMS = 3.8 \%. The average value of
$\disp\frac{V^2_{\rm true}-V^2_x}{V^2_{\rm true}}$ over $R_{\rm
  eff}<R<2R_{\rm eff}$ is 6.8 \%.

\begin{figure*}
\plotwide{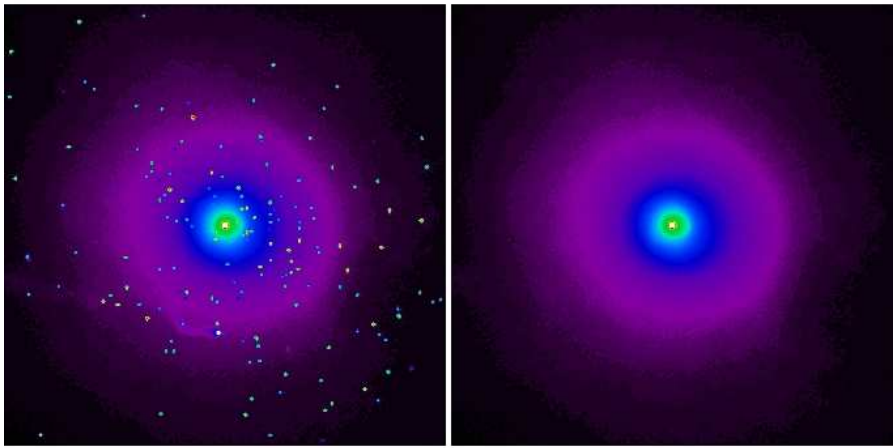}
\caption{Projected number density of hot gas in simulated
  galaxy. Left: initial density, right: density with excluded clumps.
\label{fig:diff}
}
\end{figure*}

To calculate averaged ratio of kinetic energy and thermal energy
$\disp \frac{E_{\rm kin}}{E_{\rm therm}}$ on $R_{\rm
eff}<R<2R_{\rm eff}$ one should exclude cold dense clumps since
their contribution to the kinetic energy can be significant. The
procedure to exclude clumps is described in \cite{Zhu11}.  In
brief, in each radial shell, we exclude particles with density
exceeding the median value by more that 2 standard deviations. An
example of
initial and diffuse projected densities is shown in
Figure \ref{fig:diff}. Calculated mean ratio of $\disp \frac{E_{\rm
kin}}{E_{\rm therm}}$ for diffuse component is 4.4 \%, which is
close to the bias in mass from HE.

\section{Discussion}
\label{sec:discuss}

In Table \ref{tab:table} we summarize the bias and accuracy of all methods discussed above. The sample of simulated galaxies was divided into 4 subsamples: the whole sample without exceptions (`A'), the subsample `G'(`good') for which merging and oblate galaxies observed along the rotation axis are excluded, the subsample `M' of massive galaxies with $\sigma(R_{\rm eff})>150$ $\kms$, and the subsample `MG' of massive galaxies when merging and oblate galaxies seen along the rotation axis are excluded. For estimations of the potential the bias and the RMS are nearly twice large as those for the circular speed estimations. To avoid possible confusion all values in the table are associated with $V_c$ - estimations.

\begin{table*}
\centering
\caption{Summary of the methods discussed. \label{tab:table} }
\begin{tabular}{|l|*{8}{c|}}
\hline
   & \multicolumn{2}{c}{A} &\multicolumn{2}{c}{G} &\multicolumn{2}{c}{M} & \multicolumn{2}{c}{MG} \\

   & $\overline{\Delta},$ & RMS, & $\overline{\Delta},$ & RMS, & $\overline{\Delta},$ & RMS, & $\overline{\Delta},$ & RMS, \\

   & $\%$ & $\%$ & $\%$ & $\%$ & $\%$ & $\%$ & $\%$ & $\%$ \\
\hline
full analysis, $R_{\rm sweet}$       & -1.8 & 8.6 & -1.2  & 6.8 & 0.2 & 7.7 & 0.0 & 5.4  \\
full analysis, $R_{\rm eff}$         & -2.0 & 8.6 & -2.4 & 5.9 & -0.6 & 8.5 & -1.0 & 5.1 \\
simplified analysis, $R_{\rm sweet}$ & -5.9 & 9.6 & -5.8 & 7.4 & -3.3 & 9.2 & -4.0 & 6.4 \\
simplified analysis, $R_{\rm eff}$   & -4.3 & 8.9 & -4.6 & 6.6 & -2.7 & 8.7 & -3.0 & 5.8 \\
$\Phi^{\rm true}=\kappa \cdot \Phi^{opt}+const$, eq.(\ref{eq:main})          &  3.7 & 8.7 & 2.9 & 7.1 & 1.2 & 6.7 & 1.2 & 5.1 \\
$\Phi^{\rm true}=\kappa \cdot \Phi^{opt}+const$, eq.(\ref{eq:agd_simple})          &  7.5 & 10.1 & 6.7 & 8.2 & 4.4 & 7.8 & 4.5 & 5.9 \\
aperture dispersions, $R_{\rm eff}$ & -1.4 & 10.3 & -1.5 & 9.2 & 1.1 & 9.2 & 1.0 & 7.8 \\
\hline
   & \multicolumn{4}{c|}{$R_{\rm eff}$} & \multicolumn{4}{c|}{$2R_{\rm eff}$} \\ 
   & \multicolumn{2}{c|}{$\overline{\Delta},\%$} & \multicolumn{2}{c|}{RMS,$\%$}& \multicolumn{2}{c|}{$\overline{\Delta},\%$} & \multicolumn{2}{c|}{RMS,$\%$} \\ 
X-ray                           &  \multicolumn{2}{c|}{-3.0} & \multicolumn{2}{c|}{4.4}& \multicolumn{2}{c|}{-4.0} & \multicolumn{2}{c|}{3.8}\\
\hline
\end{tabular}
\end{table*}

In case of the subsample `MG' the estimation of the circular speed at the sweet point with help of equations (\ref{eq:main}) gives the unbiased result ($\overline{\Delta_{opt}}\simeq 0 \%$) and reasonable accuracy (RMS $\simeq 5-6\%$). To test whether the unbiased average is not just a coincidence we have performed a `Jack knife' test. The resulting average for randomly chosen subsamples is less than 1$\%$. The subsample `MG' consists of 106 objects and the statistical uncertainty in this case is about $0.9 \%$.

For the subsample `M' of massive galaxies (127 objects, 26 of them ($13.3 \%$) are oblate, 3 of them ($2.4 \%$) are with ongoing merger activity) we also got almost the unbiased average ($\overline{\Delta_{opt}}=(0.2 \pm 1.2) \%$). From an observational point of view merging objects can be easily excluded while information on the `oblateness' of galaxies may not be available. If we exclude merging galaxies from the subsample `M' we get the average value of the deviation $\overline{\Delta_{opt}}=-0.7 \% $, RMS = 5.9$\%$. So the result is almost unbiased. But if run the `Jack knife' tests we get on average slightly underestimated values of the circular speed with $\left|\overline{\Delta_{opt}}\right|$ less than 1.5 $\%$.  

\begin{figure*}
\plotone{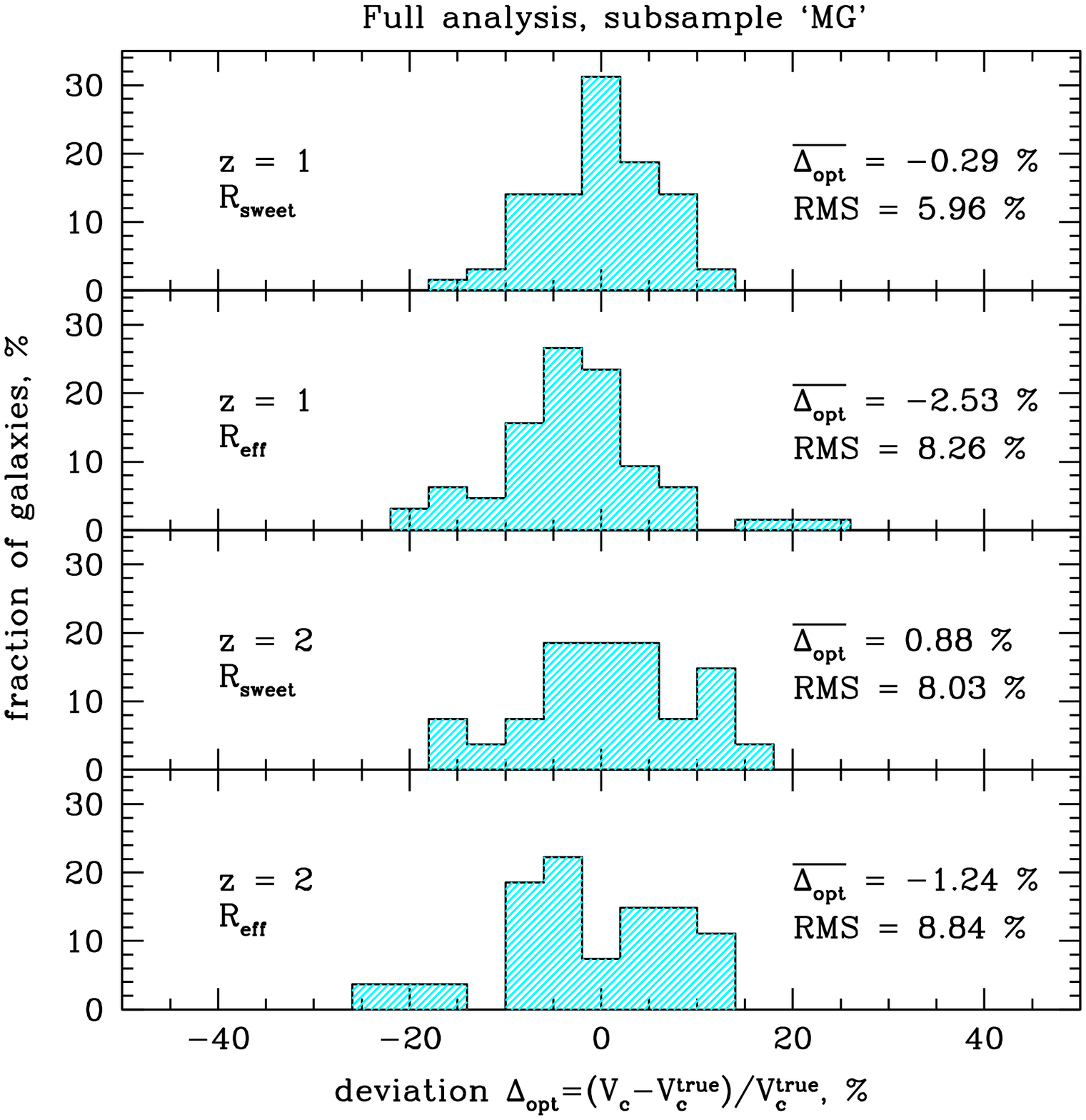}
\caption{Distribution of high-redshift galaxies from the subsample `MG' (massive galaxies with $\sigma(R_{\rm eff}) > 150$ $\kms$ when merging and oblate galaxies observed along the rotation axis are excluded) according to their circular speed deviations. Deviations are calculated at $R_{\rm sweet}$ (panels (A), (C)) and at $R_{\rm eff}$ (panels (B), (D)). 
\label{fig:z=12}
}
\end{figure*}

The method is not restricted to nearby galaxies, it also allows to recover the circular speed for high-redshift ellipticals. The circular speed estimate averaged over the subsample of massive and slowly or non-rotating simulated objests (mergers are excluded) at $z=1$ is $\overline{\Delta_{opt}}=(-0.3\pm 1.1) \%$ with RMS = 6.0 \%, at $z=2$ the average deviation is $\overline{\Delta_{opt}}=(0.9 \pm 2.2)\%$ and RMS = 8.0 \%. So the $V_c$-estimates are also almost unbiased with modest scatter of $6-8\%$ as in case of subsample `MG' at $z=0$.

While derivation of equations (\ref{eq:main}) and
  (\ref{eq:agd_simple}) is based on the assumption of the logarithmic
  form of the gravitational potential we have shown that the circular
  speed estimate at the sweet point is still reasonable even if true
  circular velocity is not flat. 

The case of slowly changing $V_c$ with radius can be illustrated by
the following example. If we assume that $V_c$ varies with radius
as a power law along with other quantities $\disp I(R)\propto
R^{-\alpha}, \sigma_p^2\propto R^{-\gamma}, \beta=\rm const$ we end up
with the following relation between $V_c$ and $\sigma_p$ (from Jeans equation):
\be
V_c^2(R)=\sigma_p^2(R)\cdot\frac{1+\alpha+\gamma-2\beta}{(1-\beta\cdot\frac{\alpha+\gamma}{1+\alpha+\gamma})}\cdot \frac{\Gamma[\frac{\alpha}{2}]\cdot \Gamma[\frac{1+\alpha+\gamma}{2}]}{\Gamma[\frac{1+\alpha}{2}]\cdot \Gamma[\frac{\alpha+\gamma}{2}]},
\label{eq:vcpow}
\ee
where $\disp \Gamma[x]$ is the gamma function. This relation is
insensitive to the anisotropy parameter $\beta$ when
$\alpha+\gamma=2$. One can hope therefore that for slopes slowly varying
with radius the sweet point will be located at the radius where this
condition is met. Substituting $\alpha+\gamma=2$ in equation (\ref{eq:vcpow})
yields the relation between  $V_c$ and $\sigma_p$ which coincides with 
 equation (\ref{eq:agd}) for isotropic orbits for $\alpha=2$.  Deviations of
$\alpha$ from 2 by 10\% cause modest $\sim 3$\% variations in $V_c$.

Simulated galaxies are of course more complicated than the above
example. For our sample we have investigated
  possible correlations between the deviation $\Delta_{opt}$ of the
  estimated $V_{c}$ from the true one and local (at $\disp R_{\rm sweet}$)
  slopes of the velocity $\disp d\ln V_{c}^{\rm true}/d\ln r$, surface
  brightness $\disp \alpha=-d\ln I(R)/d\ln R$ and velocity dispersion
  $\disp \gamma=-d\ln \sigma^2/d\ln R$ profiles. There is no obvious
  correlation between $\disp \Delta_{opt}$ and $\alpha$ or
  $\gamma$. We do see a 
  weak linear trend in $\disp
  \Delta_{opt}$ and $\disp d\ln V_{c}^{\rm true}/d\ln r$, although it is much
  smaller than the scatter in $\disp
  \Delta_{opt}$. Most of the galaxies in the sample
  $V_c(R)$ slowly declines with radius near $R_{\rm sweet}$ (see
  Figure \ref{fig:isothermal}). However, even after subtracting this trend,
  the RMS-scatter in $\disp \Delta_{opt}$ is reduced from 5.4\% to
  5.0\%, i.e. only by 0.4\%.

Comparable results are obtained using $\int \left[V_c^{\rm iso}\right]^2/r \,dr$ over $[0.5R_{\rm eff}, 3R_{\rm eff}]$ as an estimator of the gravitational potential. 

The simplified version of the analysis (equations (\ref{eq:agd_simple})) at the sweet point gives almost the same result as at the effective radius. So if one has no enough data to calculate all necessary for applying equations (\ref{eq:main}) derivatives it makes sense to derive $V_c^{\rm iso}$ from the first formula of (\ref{eq:agd_simple}) and use $V_c^{\rm iso}(R_{\rm eff})$ as an estimation of the circular speed. The quality of such approach depends on the `quality' of the sample. In case of non-interacting and almost spherical galaxies the RMS is about $7\%$ and the bias is about $(- 4 \pm 1.1) \%$. Assuming flat projected velocity dispersion profile leads to the underestimation of the circular speed. If data on the line-of-sight velocity dispersion allow to estimate the overall trend $\Delta \sigma/ \Delta R$ it may reduce the bias.

In general we can expect the sweet point to be not far from the radius  $R_2$ where $\disp -d \ln I(R) / d \ln R \simeq 2$.
Indeed for a smooth surface brightness profile which  gradually steepers with radius an integral $ \int I(R)RdR$ diverges 
at low or high limits for $\disp -d \ln I(R) / d \ln R$ greater or lower than $2$ respectively. 
Therefore one can hope that at the radius $R_2$  the contributions to the integral 
of $R<R_2$ and $R>R_2$ to be comparable and $R_2 \sim R_{\rm eff}$. Thus, $R_{\rm sweet} \sim R_{\rm eff}$. E.g. for a S\'{e}rsic model with index $n$ (\cite{2005PASA...22..118G})

$$ -\frac{d \ln I(R)}{d \ln R}\simeq 2 \left( \frac{R}{R_{\rm eff}} \right)^{1/n}  $$

and it can be easily seen that the sweet point for the circular speed estimation is of order of the effective radius. Moreover, 
as it was shown in \cite{2010MNRAS.404.1165C} (Table~4) for S\'{e}rsic models the stellar anisotropy is close to minimal at about $0.5 R_{\rm eff}$ and this radius can be used as the sweet point for the circular speed determination. 

We have tested the statement that $R_{\rm sweet} \sim R_{2} \sim R_{\rm eff}$ on the sample of the simulated objects. If the slope of the surface brightness profile is close to $-2$ over some range of radii or $\disp \alpha=-d \ln I(R) / d \ln R$ is not monotonic then there is an ambiguity in selecting $R_{\rm sweet}$ and $R_{2}$. To avoid this ambiguity we have smoothed $I(R)$ and $\sigma(R)$ using the width of the window function $\Delta_{I}=\Delta_{\sigma}=1.0$. As a result $\alpha(R)$ has become monotonic for majority of objects and newly determined $\tilde{R}_{\rm sweet}$, $\tilde{R}_{2}$ follow the relationship $\tilde{R}_{\rm sweet} \sim \tilde{R}_{2} \sim R_{\rm eff}$. However, a significant smoothing of data leads to a bias in estimating the circular speed $\disp \overline{\Delta_{opt}} \simeq -2 \%$ at both $\tilde{R}_{\rm sweet}$ and $\tilde{R}_{2}$.

\section{Conclusions}
\label{sec:conc}

Being an important issue, the total mass estimation for elliptical
galaxies is often quite difficult, especially for galaxies at high
redshift. We used a large sample of cosmological zoom simulations of
individual galaxies to test a simple and robust procedure (see
equations (\ref{eq:main}), (\ref{eq:agd_simple})) based on the surface brightness and velocity dispersion
profiles to estimate the circular speed and therefore the total mass
of a massive galaxy. The method is very simple and it does not require any
assumptions on the stellar anisotropy profile. For massive ellipticals without
significant rotation at redshifts $z=0-2$ it gives an unbiased
estimate of the circular speed (the bias $\Delta_{opt}(R_{\rm sweet})$
is less than 1\%) with 5-6\% scatter. Therefore this method
is suitable for the analyze of large samples of galaxies with limited
observational data at low and high redshifts. The method works best
for the most massive ellipticals ($\sigma(R_{\rm eff})>200$ $\kms$), which
in the present simulations have almost isothermal circular velocity
profiles over broad range of radii.

The method should be applied with caution to merging galaxies where
the circular speed can be significantly overestimated. For rotating
galaxies seen along the rotation axis the procedure gives
substantially underestimated $V_c$.

The best estimate of the circular speed is obtained at a sweet point
$R_{\rm sweet}$ where the sensitivity of the recovered circular speed
to the stellar anisotropy is expected to be minimal (see section \ref{sec:method}).
The $R_{\rm sweet}$ is expected to be not far from the projected
radius where the surface brightness declines approximately as
$I\propto R^{-2}$. This radius is in turn close (within factor of 2)
to the effective radius $R_{\rm eff}$ of the galaxy. Our tests have
shown that the accuracy (RMS scatter) of the circular speed estimates at $0.5-2~
R_{\rm eff}$ is $5-7\%$ for most massive ellipticals. 

An even simpler method - based on the aperture velocity dispersion
(equations (\ref{eq:vcap}), (\ref{eq:aperture})) - is found to be less accurate, although the results are
still reasonable. For example, for massive galaxies without
significant rotation the sample averaged deviation of the circular
speed at the effective radius is $\Delta_{opt}(R_{\rm eff})=(1.0 \pm 1.3) \%$
with RMS $\simeq 8\%$. Other flavors of the circular speed estimates
are described in Section \ref{subsec:aperture}.

Using the same simulated set we have also tested the accuracy of the
circular speed estimate from the hydrostatic equilibrium equation for
the hot gas in massive ellipticals. We found a negative bias at the
level of $3-4 \%$ and the scatter of $\simeq 5\%$. The presence of bias
is caused by the residual gas motions.

Given the simplicity of the described method (Section \ref{sec:method}), the low bias
and modest scatter in the recovered value of the circular speed, it is
suitable for the analysis of large samples of massive elliptical
galaxies at low as well as high reshifts.

\section{Acknowledgments} 
We are grateful to the referee for very useful comments and suggestions.
NL is grateful to the International Max Planck Research School on Astrophysics (IMPRS) for financial support. TN acknowledges support from the DFG Excellence Cluster "Origin and Structure of the Universe". The work was supported in part by the Division of Physical Sciences of the RAS (the program ``Extended objects in the Universe'', OFN-16).

\label{lastpage}
\end{document}